\documentstyle[12pt]{article}
\input epsf
\begin{document}
\pagestyle{plain}
\pagenumbering{arabic}
\date{January 20, 1997}
\baselineskip=12pt
\begin{center}
\begin{large}
{\bf Diffusive Evolution of Stable and Metastable Phases I:
Local Dynamics of Interfaces}\\
\end{large}
\medskip
{\em R. M. L. Evans  \\
     M. E. Cates} \\
\medskip
Department of Physics and Astronomy \\
The University of Edinburgh \\
JCMB King's Buildings\\
Mayfield Road, Edinburgh EH9 3JZ, U.K.\\
e-mail: r.m.l.evans@ed.ac.uk \\
\bigskip
\medskip
{\bf Abstract}
\end{center}

\begin{quote}
We find analytical solutions to the Cahn-Hilliard equation for the dynamics of
an interface in a system with a conserved order parameter (Model B). We show
that, although steady-state solutions of Model B are unphysical in the far-field,
they shed light on the local dynamics of an interface. Exact solutions are given
for a particular class of order-parameter potentials, and an expandable integral
equation is derived for the general case. As well as revealing some
generic properties of interfaces moving under condensation or evaporation, the
formalism is used to investigate two distinct modes of interface propagation in
systems with a metastable potential well. Given a sufficient transient increase
in the flux of material onto a condensation nucleus, the normal motion of the
interface can be disrupted by interfacial unbinding, leading to growth of a
macroscopic amount of a metastable phase.
\end{quote}



\baselineskip=16pt

\section{Introduction}

  The kinetics of phase ordering is a central topic in nonequilibrium
statistical physics. Much of our understanding is based on theories
that describe one or more slowly-varying density (or order parameter) variables,
governed by a local Langevin equation \cite{Chaikin1}. In general,
the density variable(s) evolve(s) systematically in response to a
driving force, which is a derivative of the underlying free energy
functional, with some mobility  (characterized by the Onsager matrix). On top of
this are noise terms whose magnitude is fixed by requiring that the Boltzmann
distribution is a stationary state of the dynamics.  The nature of the
Onsager mobility depends on the kind of ordering involved; specifically
we must distinguish conserved order parameters from nonconserved ones.
In the conserved case, the density in some region can change only by
diffusive transport across its boundary; its time derivative is
therefore the divergence of a current. This is not the case for
nonconserved order parameters, which can change locally in direct
response to the driving force.

The low temperature (noise-free) limit is usually considered
appropriate for the study of phase ordering kinetics, in which a system
is prepared far from equilibrium and then allowed to evolve. For
example, a uniform high-temperature phase can be quenched into a region
where it is either locally or globally unstable with respect to
separation into two macroscopic phases. Local instability leads to
spinodal decomposition \cite{Gunton83}; if the system is locally stable, phase
separation proceeds by a nucleation and growth mechanism \cite{Langer95}. In
either case, the governing equation for phase separation of a conserved
scalar order parameter is the Cahn-Hilliard equation \cite{Cahn58}:
\begin{equation}
\label{CHintro}
 \frac{\partial\rho}{\partial t} = \mbox{\boldmath $\nabla$}\cdot \left(\Gamma
  \mbox{\boldmath $\nabla$} \left\{
  \frac{df}{d\rho}-K\nabla^2\rho \right\} \right).
\end{equation}
As explained in Section \ref{CH}, $K$ is the square-gradient
coefficient in a free energy expansion, which treats the order parameter,
$\rho$, as slowly varying; $\Gamma$ is the mobility.  Note that in principle
Cahn-Hilliard theory can accommodate an arbitrary form of $f(\rho)$, which is
the free energy density for a homogeneous state. (In particular, it does not
assume that $f$ is a polynomial in the order parameter $\rho$, as would be
assumed in the time dependent Landau-Ginzburg theory of dynamics close to a
critical point \cite{Chaikin2}.) Indeed, the approach should be qualitatively
applicable even if $f(\rho)$ consists of the lower envelope of several unrelated
functions representing phases of different symmetry. The free energy near a
liquid-solid transition is of this form, for example, with $\rho$ the material
density. This assumes only that, whatever other order parameters distinguish
the various phases (such as crystallinity), these can relax quickly and
hence that the rate-limiting process for the phase ordering is transport of
$\rho$. It is conventional, in equation \ref{CHintro}, to treat $K$
as a constant (independent of $\rho$). We do this in what follows, although
it might be a dangerous assumption when $f$ is a composite function as
just described. With this caveat, equation \ref{CHintro} will be relevant at
long times and large distances if the other order parameters are nonconserved.

In this paper, we therefore consider the phase ordering problem for
relatively general forms of $f(\rho)$, where $\rho$ is a concentration
variable. We assume this is the {\em only} conserved order parameter, thereby
ruling out systems with significant concentration deviations in more
than one species, and also ruling out consideration of heat transport.
This latter restriction might be severe in (say) metallurgical
applications, but not for soft condensed matter systems (such as
colloidal suspensions) which are our main interest. Indeed, for many
such systems the latent heats of phase changes are entirely negligible
\cite{Lowen94}.

Our work is motivated by the desire to understand better the role of
metastable phases in the kinetics of phase separation.  That such a
role exists has been long acknowledged: for example, the ``Ostwald rule
of stages" \cite{Ostwald} asserts that a system will progress from an unstable
to a stable state, not directly, but by a sequence of steps through any
intervening metastable states that may be present. In the area of
metallurgy, there is an extensive folklore on the subject \cite{DeHoff93}.
Here we
aim at a more fundamental understanding, based on a direct analysis of
the Cahn-Hilliard problem. For the most part, we work in one space
dimension.

Specifically, we shall focus on {\em steady-state solutions} of the
Cahn-Hilliard
equation, in which interfaces between phases move with constant velocities. This
approach appears at first paradoxical, since the diffusive nature of the
transport rules out true constant-velocity solutions when conserved order
parameters are present. (Indeed the basic scaling of lengths in diffusive
transport is with $t^{1/2}$ whereas constant velocities would imply linear
scalings.) However, as we discuss later, significant physical insights can be
gained by viewing the interfacial motion as a quasi-steady process. A broadly
comparable analysis, for nonconserved dynamics (where true steady-state motion
is possible) has been given recently by Bechhoefer et al.\
\cite{Bechhoefer91,Celestini94,Tuckerman92}.
These authors showed that under sufficient supercooling (or,
equivalently in a ferromagnetic system, sufficient applied field) the
interface between two stable coexisting phases could become dynamically
unstable toward ``splitting". The splitting instability results in a
macroscopically thick slab of a metastable phase appearing between
the two stable phases, which can then grow.

One of the main questions we address here, and in a companion paper
\cite{Evans97b} is whether the same scenario is possible for the conserved
order parameter case. In this paper we show that, although there is no
mathematically direct analogue of the splitting instability found by
Bechhoefer et al., a sufficient transient flux from the less dense to
the more dense stable phase can indeed cause interfacial splitting. We
also argue that the split mode will be maintained so long as the
supersaturation of the less dense phase is sufficiently large. In the
companion paper \cite{Evans97b} we study the long time limit in which the
interfaces become sharp on the scale of their separation, and give a more
detailed discussion of the critical supersaturation required to sustain
the split mode at long times. That paper also contains a discussion of
experimental evidence, involving colloid-polymer mixtures \cite{Poon95},
which suggests that the onset of the split mode might be connected with
the observation of arrested crystallization, beyond a threshold of
supersaturation, in the transition from a colloidal fluid to a
colloidal crystal. A brief account of these ideas is given in \cite{Evanslet}.

The rest of this paper is organized as follows. In Section \ref{CH} we
recall the Cahn-Hilliard equation and discuss the conditions under
which a one-dimensional treatment should suffice. In Section
\ref{exact} we formulate the quasi steady-state form of the
Cahn-Hilliard equation, paying careful attention to the boundary
conditions that are required to make the solution
physically meaningful. An exact solution is described for a piecewise
quadratic potential $f(\rho)$, with piecewise constant
mobility. In Section \ref{properties} we describe in more detail the
properties of the solution, focusing on the case where $f(\rho)$ shows
a metastable minimum of intermediate density. We argue that there is no
critical velocity above which the interface between stable phases
ceases to have a steady-state solution (in contrast to the nonconserved
case) and in Section \ref{stability} we show explicitly that such an
``unsplit" mode of interface motion is linearly stable. In Section
\ref{splitdyn}, however, we show that a split interface, should one arise, can
also be dynamically stable under appropriate conditions. We discuss
qualitatively the nature of the (large) perturbation required to cause
splitting.  Section \ref{conclusion} summarizes our conclusions.
Appendix A provides some details of the exact solution for the piecewise
quadratic case, whereas in Appendix B we derive an exact integral representation
of the steady-state solution for general potentials, thereby confirming
and extending some of the earlier results.

\section{The Cahn-Hilliard Equation}
\label{CH}

  Consider a system characterized by one conserved, scalar order
parameter, such as local mass density, in a part of the phase
diagram where two-phase coexistence is the equilibrium state. If the
system is far from criticality, and is initially out of equilibrium,
then its evolution towards equilibrium obeys Model B, described by equation
\ref{CHintro} (the Cahn-Hilliard equation), which is derived as
follows. Let the free energy of the system be a functional $F[\rho]$ of
the order parameter $\rho(\mbox{\boldmath $x$})$. Then the chemical potential is
defined by the functional derivative,
\begin{displaymath}
  \mu = \frac{\delta F}{\delta\rho}
\end{displaymath}
($\mu$ can thereby depend on gradients of
$\rho$, as well as
$\rho$ itself). Currents in Model B are induced by gradients of the
chemical potential:
\begin{displaymath}
  \mbox{\boldmath $j$} = -\Gamma \mbox{\boldmath $\nabla$} \mu
\end{displaymath}
where the constant of proportionality, $\Gamma$ is the Onsager
mobility, which may be a function of $\rho$. Since the
order parameter is conserved, its time-derivative is given by a
continuity equation,
\begin{displaymath}
  \frac{\partial\rho}{\partial t} =
   -\mbox{\boldmath $\nabla$} \cdot \mbox{\boldmath $j$}.
\end{displaymath}

 Let the free energy functional be of the form
\begin{displaymath}
  F[\rho] = \int d^d x \left\{ f(\rho) + \mbox{$\frac{1}{2}$} K
  (\mbox{\boldmath $\nabla$}\rho)^2 \right\}.
\end{displaymath}
The form of the bulk free energy density (or `order-parameter
potential') $f$, and the value of $K$, are system-dependent. Let the
system in question be initially homogeneous at a non-equilibrium value
of $\rho$, between two minima in $f$; {\em i.e.\ }within a two-phase
coexistence region of the phase diagram. For definiteness, let
$\rho$ be initially close to the low-density minimum in $f$. Small
fluctuations initially induce the evolution of $\rho$ towards
equilibrium. Early stages of the evolution proceed by nucleation, if $f$
is convex at the given value of $\rho$, or by spinodal decomposition, if
concave. This stage of the dynamics is not addressed here. By whatever
process, domain walls (interfaces) soon form. If surface tensions are neglected
(legitimate when typical interfacial radii of
curvature are large, which we assume), subsequent motion of a wall is driven
by diffusion from the far-field. The local profile of the wall changes on
a shorter time scale than the far-field gradients which determine the
flux onto the wall, simply because of the difference in length
scales: while typical distances between interfaces are proportional to
$t^{1/2}$, the characteristic width an interface remains of the order of
$\sqrt{K/(d^2f/d\rho^2)}$ at all times. Therefore, although typical inter-wall
distances vary with time, intra-wall dynamics (concerning the {\em local}
density profile of an interface) soon become approximately steady-state, with
a quasi-constant input flux.

These {\em local} interface dynamics are concerned with the movement of
a $(d-1)$-dimensional wall in a $d$-dimensional space, translating
normal to itself. Hence, if curvature and surface-tension effects are ignored,
the problem becomes one-dimensional; the Cahn-Hilliard equation then reduces to
\begin{displaymath}
  \frac{\partial\rho}{\partial t} = \frac{\partial}{\partial x}
   \left(\Gamma(\rho)\frac{\partial}{\partial x}
   \left( \frac{df}{d\rho}-K\frac{\partial^2\rho}{\partial x^2}
   \right) \right).
\end{displaymath}
In the next Section, we show how to find steady-state solutions of this
equation for a piecewise-quadratic potential.

\section{Exact Steady-State Solution}
\label{exact}

Let us transform to a frame moving at velocity $v$, in which the
position coordinate is
\begin{displaymath}
  s \equiv x-vt,
\end{displaymath}
and introduce the notation $f_{n}(\rho)\equiv(d/d\rho)^nf(\rho)$.
Demanding that the time derivative vanishes in this frame, we find that
a steady-state solution $\rho(s)$ to the Cahn-Hilliard equation,
travelling at velocity $v$ satisfies
\begin{equation}
\label{steady}
  \rho''''-\frac{f_{2}}{K}\rho''-\frac{v}{K\Gamma}\rho'
    = \frac{f_{3}}{K}\rho'^2 +\frac{1}{\Gamma}\frac{d\Gamma}{d\rho}\left(
    \frac{f_{2}}{K}\rho'-\rho'''\right) \rho'
\end{equation}
which is a third-order ordinary differential equation in
$\rho'$($\equiv d\rho/ds$). If $\Gamma$ and $f_{2}$ are both
independent of $\rho$ then the right-hand side vanishes and the
left-hand side becomes linear and homogeneous in $\rho'$. So the problem
is (at least) piecewise-soluble for a piecewise-quadratic potential
$f(\rho)$, with piecewise-constant mobility $\Gamma$. Such a model is
actually fairly versatile, and we therefore explore it in detail.

  At discontinuities in our piecewise-constant $f_{2}$ and/or $\Gamma$, the
solutions for the separate pieces must respect certain matching conditions.
Specifically, the current must be continuous so that $\nabla\cdot j$ remains
finite; the chemical potential must be continuous to avoid infinite
currents; and continuity of the gradient of $\rho$ is required to avoid
infinities in the chemical potential. These three conditions are
sufficient to fix the constants of integration for the above
third-order equation in $\rho'$. Subsequent integration to find $\rho$
gives rise to an additional arbitrary constant, which is fixed by
demanding continuity of $\rho$ itself. The full solution is given in
detail in Appendix \ref{solution}. For each piece of the
potential, this solution has the form
\begin{equation}
\label{soln}
  \rho'(s)=\sum_{j=1}^{3} A_{j} e^{\omega_{j}s}
\end{equation}
where the constants $A_{j}$ and $\omega_{j}$ depend on $f_{2}$, $K$,
$\Gamma$ and $v$.

  In addition to the matching conditions (which fix some of the constants
arising in equation \ref{soln} from the division of $f(\rho)$ and
$\Gamma(\rho)$ into pieces) boundary conditions are required, to select a
specific solution to the differential equation. The correct choice and
interpretation of these boundary conditions are not trivial; we
discuss them carefully before proceeding further. The case of
interest
\begin{figure}[h]
  \epsfxsize=10cm
  \begin{center}
  \leavevmode\epsffile{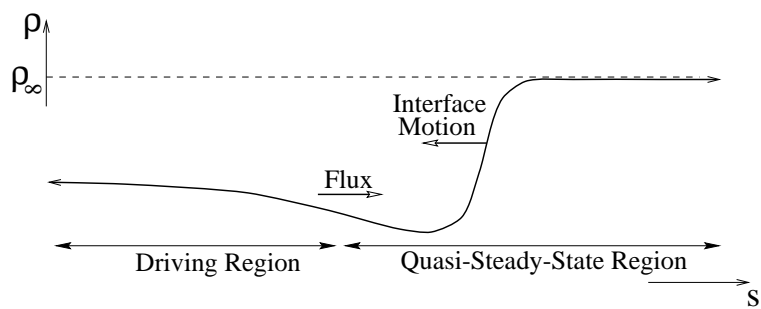}
  \caption{Schematic diagram of the system. Density
   $\rho$ is plotted against distance $s$. On the right of the figure, a
   high density domain has formed, for which
   $\rho\rightarrow\rho_{\infty}$ as $s\rightarrow+\infty$. On the low
   density side of the interface, the region for which the
   quasi-steady-state approximation holds meets the `driving region',
   which becomes depleted as material condenses onto the interface.}
  \label{schematic}
  \end{center}
\end{figure}
(depicted in figure \ref{schematic}) is where a region of the
high-density phase has formed, and is growing by condensation from the
supersaturated low-density phase. (Note that, throughout this paper, the
high-density phase is depicted on the right of the diagrams, and therefore grows
by leftward motion of the interface. The opposite convention is adopted in
Refs.\ \cite{Evans97b} and \cite{Evanslet}.) The interface is to be modelled in
isolation, so the high-density phase is semi-infinite. Two boundary
conditions arise from this: that the density (and hence chemical
potential) asymptotes to a constant value, $\rho_{\infty}$ as
$s\rightarrow+\infty$ (say), and that the flux asymptotes to zero in
this limit. We may now either regard $v$ in equation \ref{steady} as a
given constant and then, from integration, deduce the conditions at the
other boundary, or $v$ may be seen as an eigenvalue which is set by
further boundary conditions. In any case, at the second boundary a flux
is required (to induce motion), and this implies a gradient in the
chemical potential. Hence it does not make sense to put this boundary
at $s\rightarrow-\infty$, as $\mu$ would be infinite here; instead the
left-hand boundary must be at some finite position. This raises two
possible worries: that the thermodynamic limit cannot be taken since
the model is of a finite system, and that steady-state solutions cannot
be found in a finite system. The interpretation which resolves these
difficulties is as follows. The left-hand boundary is at a finite
distance from the interface, and moves with the interface. It is not in
fact the edge of the system, but simply the point at which the
behaviour ceases to be quasi-steady-state. The part of the system to
the left of this boundary, which does not solve the steady-state
equation, may be referred to as the `driving region', since it is
responsible for supplying the quasi-constant flux and chemical
potential to the propagating interface, by non-steady-state diffusive
depletion of material. In summary, a specific solution to the
steady-state equation is fixed by the asymptotic value of $\rho$ in the
limit $s\rightarrow+\infty$ and by the values of the chemical potential
and flux at some (rather ill-defined) position to the left of the interface,
where the steady-state region meets the driving region.

\section{Properties of the Solution}
\label{properties}

One important qualitative observation, noticeable on graphs (such as those
discussed in section \ref{form}) of the exact steady-state solution calculated
in Appendix \ref{solution}, is that the characteristic width of the interface
always {\em decreases} as the speed of condensation (or equivalently the
incident flux) {\em increases}. In Appendix \ref{generic}, it will be shown,
to first order in $v$, that this is a generic result. A second qualitative
result, found below (Section \ref{nosplit}), is that steady-state solutions
exist at all velocities $v$. This holds even when an intermediate metastable
well is present in the order parameter potential; accordingly (and in contrast
to the case of a nonconserved order parameter \cite{Bechhoefer91}) there is no
critical velocity above which the interface {\em must} split. Both of these
results have implications for the formation of metastable phases.

\subsection{Form of the Interfacial Profile}
\label{form}
\begin{figure}[h]
  \epsfysize=12cm
  \begin{center}
  \leavevmode\epsffile{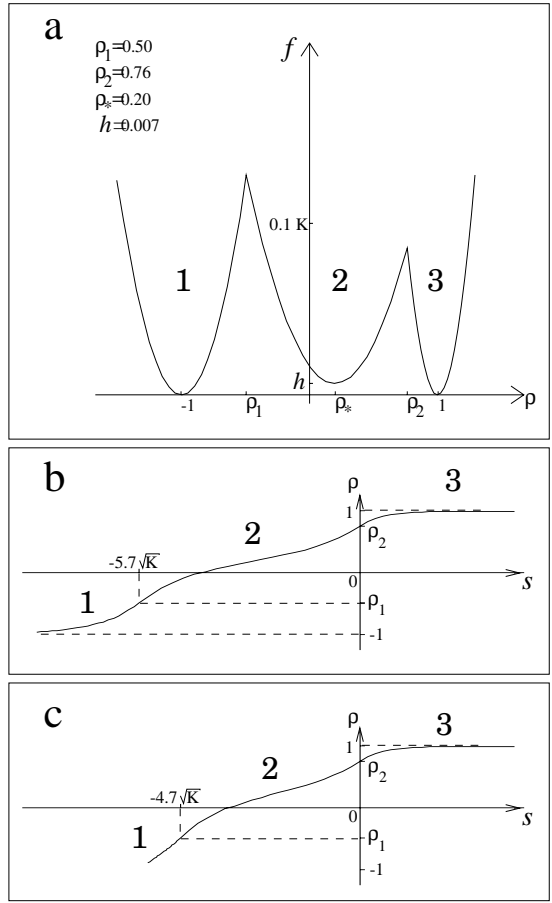}
  \caption{(a) A piecewise-quadratic three-well potential $f(\rho)$,
   for which all relevant parameters are given. Well number 2 is
   metastable. (b) The density profile of an equilibrium interface for the
   above potential. Regions are labelled in which the density corresponds
   to wells 1, 2 and 3 of the potential. (c) Density profile of an
   interface moving at velocity $-0.02K\Gamma$ for the same potential, and
   constant mobility $\Gamma$. Notice that region 2 is narrower than in
   the equilibrium case.}
  \label{3wells}
  \end{center}
\end{figure}

Before discussing splitting, we show some typical numerical results for a
steadily moving interface, in a system where a metastable phase is possible.
Consider a system in which $f(\rho)$ contains a metastable well at a density
between that of the growing high-density phase, and the
supersaturated low-density phase. A piecewise-parabolic form for such a
potential is shown in figure \ref{3wells}a. (Metastability requires
that the middle well is
{\em above} the common tangent to the other two wells.) Let the low,
intermediate and high density wells be referred to as 1, 2 and 3 respectively.
We may define the order parameter $\rho$ (chosen as a scaled, relative density)
to be $\pm 1$ at the minima corresponding to phases 1 and 3, and take their
free energy densities $f(-1),f(+1)$ to be equal.
(This choice involves no loss of
generality, since, as is well-known \cite{Bray94}, adding a linear term
to the free energy density ($f\rightarrow f+a\rho+b$) has no effect on
the solutions of the Cahn-Hilliard equation.) Steady-state interface
profiles for this system are given in figure \ref{3wells}b for $v=0$
({\em i.e. }the equilibrium wall) and (in fig. \ref{3wells}c) for $v=-0.02
K\Gamma$. In each case the mobility $\Gamma$ is constant throughout the
system.  Negative-$v$ solutions
are of greatest interest, since they describe the condensation of
material from a supersaturated region onto a growing domain, and are
therefore central to phase-ordering dynamics.  Notice that both
solutions shown have an inflection at the density of the metastable
phase 2. Without forming a {\em macroscopic} amount of the metastable phase, the
interface takes advantage of the local minimum in free energy by having
extra material at this density. Region 2 is noticeably narrower in figure
\ref{3wells}c (for condensation) than in figure
\ref{3wells}b (the equilibrium profile).

\subsection{Existence of Solutions for an Unsplit Interface for All $v$}
\label{nosplit}
We are interested in whether such a steady-state interface might split into
two parts, the 1-2 part of the interface propagating faster than the
2-3 part, analogously to the `dynamic splitting instability'
\cite{Bechhoefer91} which can arise in the dynamics of a non-conserved
order parameter in a three-welled potential. In the non-conserved case, a
critical velocity exists, above which there exists no steady-state
solution for the propagation of a 1-3 interface; instead a macroscopic
amount of the metastable phase 2 {\em must} be created between a pair of
moving (1-2 and 2-3) interfaces.

In comparing the conserved and
non-conserved dynamics however, an important distinction should be borne in
mind. In the non-conserved case, the velocity of each interface is controlled by
an external field, which adds a linear term to the potential. (Indeed, to obtain
a dynamic splitting instability in the non-conserved case, the field
must cause the potential in the middle well to fall {\em below} that of one
of the others.) In the conserved case, however, linear terms in the
potential are
irrelevant; instead,
the velocity is controlled by the boundary conditions.
We now present an argument showing that the unsplit propagation mode exists for
all velocities in this case. (The argument is {\em not} limited to the case of
piecewise quadratic potentials.)

First, note that equation \ref{steady} may in principle be integrated spatially,
from right to left, for a given $v$ and $\rho_{\infty}$, to find the value
of $\rho$ at any point. This could fail to produce a solution for a
three-well potential, only if the resultant profile $\rho(s)$ fails to span
all three wells due to the presence of a minimum in the function $\rho(s)$.
This would occur whenever the given value of $v$ was above the critical
value. However, such turning points in $\rho(s)$ {\em do not arise} in the
steady-state solutions. This follows from the expression for the chemical
potential
\begin{displaymath}
  \mu = \frac{df(\rho)}{d\rho}-K\frac{\partial^2\rho}{\partial x^2}\;.
\end{displaymath}
To see why, consider first the equilibrium interface profile, for which $\mu$ is
a constant. Clearly, this spans all three wells. Any solution of negative $v$
(describing condensation) must have a higher chemical potential than
the equilibrium value, at any given point on the interface where
$\rho<\rho_{\infty}$, because there is a steady flux onto the high-density
phase. Hence, for any given value of $\rho$, it follows from the above
expression
for $\mu$ that the curvature of $\rho(s)$ must be {\em more negative} than for
the equilibrium profile. So no minimum exists in $\rho(s)$. This argument also
holds if
$\rho_{\infty}$ is greater than the equilibrium value, since there is
still no minimum to the $v=0$ solution in this case. The argument may
even be extended to the case where $\rho_{\infty}$ is below the
equilibrium value so that the $v=0$ solution exhibits a minimum. So
long as this static solution spans all three wells, negative $v$
solutions with the same asymptote must also do so, since their
curvature at any given value of $\rho$ is more negative. Hence there is
no critical velocity for condensation.

\section{Stability of the Solutions}
\label{stability}

Having established the {\em existence} of solutions corresponding to unsplit
interfacial propagation at all velocities $v$, we now show that these
solutions are stable against linear perturbations. (The argument below {\em is}
restricted to piecewise quadratic $f(\rho)$.)

Let the field
$\rho(x,t)$ obey the full Cahn-Hilliard equation of motion, and be written as
\begin {displaymath}
  \rho(x,t)=\rho_{0}(x,t)+\varepsilon(x,t)
\end{displaymath}
where $\rho_{0}(x,t)$ is a solution of the steady-state equation, and
$\varepsilon$ is initially small. Differentiating with respect to time
gives
\begin{displaymath}
  \frac{\partial\varepsilon}{\partial t} =
   \frac{\partial}{\partial x}\left(\Gamma
   \left.\frac{\partial\mu}{\partial x}\right|_{\rho_{0}+\varepsilon}\right)
   - \left.\frac{\partial}{\partial x}\left(\Gamma\frac{\partial\mu}{\partial x}
   \right|_{\rho_{0}}\right)
\end{displaymath}
where
\begin{displaymath}
  \mu|_{\rho_{0}+\varepsilon} = \mu|_{\rho_{0}} + \varepsilon f_{2}(\rho_{0})
   - K \frac{\partial^2\varepsilon}{\partial x^2} + {\cal O}(\varepsilon^2).
\end{displaymath}
For the exact solutions calculated in Appendix \ref{solution}, $f_{2}$
and $\Gamma$ are both piecewise constant. It follows that, on any piece
of this solution, a small perturbation $\varepsilon$ about the solution
obeys the linearized equation of motion
\begin{displaymath}
  \frac{\partial\varepsilon}{\partial t} =
   \Gamma f_{2} \frac{\partial^2\varepsilon}{\partial x^2}
   - \Gamma K \frac{\partial^4\varepsilon}{\partial x^4}.
\end{displaymath}
In Fourier space, writing
\begin{displaymath}
  \varepsilon(x,t) = \int_{-\infty}^{\infty} \tilde{\varepsilon}(q,t)
   e^{iqx}\,dq
\end{displaymath}
the equation of motion becomes
\begin{displaymath}
  \frac{\partial\tilde{\varepsilon}(q,t)}{\partial t} = -q^2\,(\Gamma f_{2}
   + \Gamma K q^2)\, \tilde{\varepsilon}(q,t).
\end{displaymath}
In any well in the potential, the coefficient $-q^2(\Gamma f_{2} +
\Gamma K q^2)$ is negative for all $q$. So all Fourier modes of a small
perturbation, about any piece of the solution in a quadratic well, decay
exponentially with time. Also, since the pieces of the solution must
always obey the matching conditions at cusps in the potential, the
solution as a whole is stable, when all continuous parts of the
potential are convex (as is the case, for instance, in the potential of
figure \ref{3wells}a). Furthermore, in any concave part of the potential,
solutions are stable with respect to perturbations of shorter
wavelength than $2\pi\sqrt{K/(-f_{2})}$. Hence, so long as the spatial
distance over which the interfacial profile spans a concave part of the
potential is less than $2\pi\sqrt{K/(-f_{2})}$, the solution will be
stable. It is obviously true that the equilibrium interface
profile satisfies this linear stability criterion. As stated previously, we
have observed solutions, for interfaces moving due to condensation, to be
{\em narrower} than the equilibrium interface (and proved it to first first
order in $v$ in Appendix \ref{generic}); hence they too are linearly stable.

\begin{figure}
  \epsfxsize=11cm
  \begin{center}
  \leavevmode\epsffile{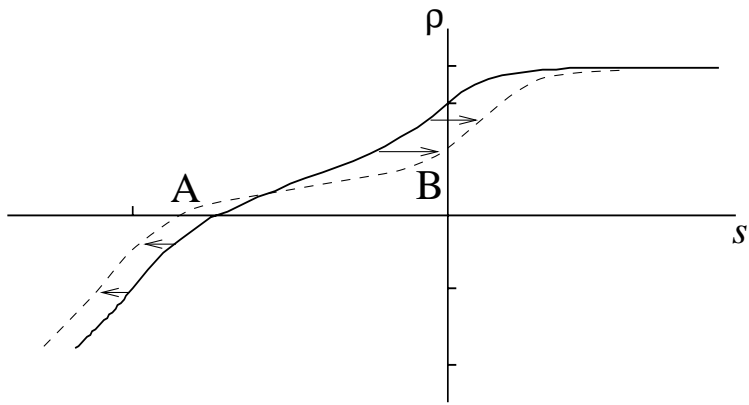}
  \caption{Density profile of a steadily moving
   interface (solid line) and the result of a perturbation in the
   direction of the arrows (dashed line) which has increased the width of
   the region occupying the metastable middle well of the potential. The
   result is a more negative curvature in the position labelled A, and
   more positive curvature at B.}
  \label{perturb}
  \end{center}
\end{figure}

Although this completes the argument, it is
useful to have a conceptual picture of the mechanisms giving rise
to this stability. Consider once more the profile in figure
\ref{3wells}c of a moving interface for a potential with a metastable
middle well. A perturbation such as that shown by the dashed line in
figure \ref{perturb} may be considered as a small change in $s$ at
constant
$\rho$, rather than {\em vice versa}. The perturbation shown is tending
to separate the \mbox{1--2} part of the interface from the \mbox{2--3}
part, thus widening the metastable region. Notice that this makes the
curvature more negative on the part of the interface labelled `A', and
more positive at `B'. So the chemical potential is increased at A and
reduced at B, thus enhancing the flux onto the \mbox{2--3} part of the
wall. So the \mbox{2--3} part of the wall will catch up with the
leading \mbox{1--2} part, and steady-state motion will be restored.
This {\em negative feedback} mechanism is responsible for the linear
stability of an unsplit interface.

  It is worth noting, in addition,
that numerical solutions of the Cahn-Hilliard equation have been performed,
using a three-well potential, and have confirmed that \mbox{1--3} interfaces
may propagate stably, even at large values of $v$.

\section{Local Dynamics of a Split Interface}
\label{splitdyn}

\begin{figure}
  \epsfxsize=12cm
  \begin{center}
  \leavevmode\epsffile{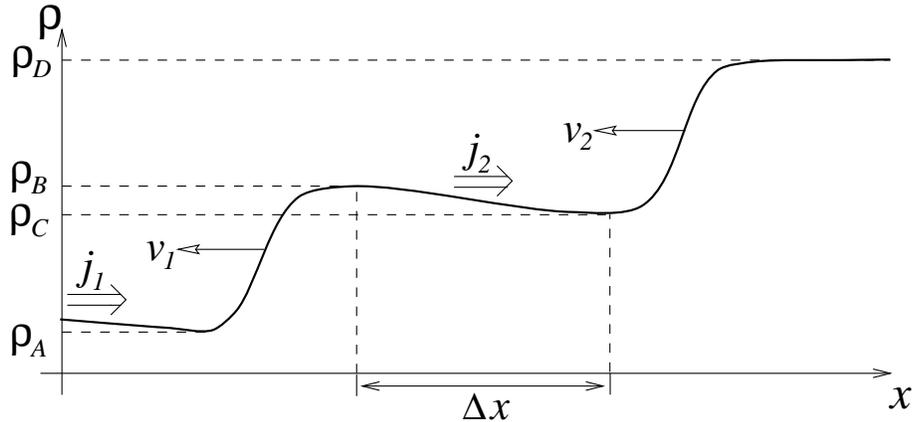}
  \caption{Density profile of a system in which \mbox{1--2} and
   \mbox{2--3} interfaces are separated by a distance $\Delta x$. The
   interfaces are travelling with velocities $v_{1}$ and  $v_{2}$
   respectively. Densities $\rho_{A}$, $\rho_{B}$, $\rho_{C}$ and
   $\rho_{D}$ and fluxes $j_{1}$ and $j_{2}$ are also defined in the
   figure.}
  \label{split}
  \end{center}
\end{figure}

  It was shown above that steady motion of a \mbox{1--3} interface,
spanning the intermediate metastable well, is linearly stable. Hence a
\mbox{1--3} interface, once formed, continues to propagate in the absence of
{\em large} perturbations. Such perturbations may however arise, especially in
the early stages of interface formation (including the dynamics prior to the
time at which the Cahn-Hilliard equation becomes a good approximation).
Therefore, let us consider the situation whereby a large slab of metastable
phase 2 has formed, by whatever mechanism, so that the \mbox{1--2} and
\mbox{2--3} interfaces are separated by a distance large compared with
the scale $\sqrt{K/f_2}$ set by the curvature term in the free energy. Such a
situation is depicted in figure \ref{split}, in which various
quantities are defined: the interface separation $\Delta x$, the two
interface velocities $v_{1}$ and $v_{2}$, the fluxes into the
\mbox{1--2} interface, $j_{1}$, and between the interfaces, $j_{2}$
(both of which are taken to be approximately constant over the spatial
regions of interest), and four densities $\rho_{A}$, $\rho_{B}$,
$\rho_{C}$ and $\rho_{D}$. Let us
introduce a further approximation, as follows. We assume that the interfaces are
moving sufficiently slowly that the densities $(\rho_{A},\rho_{B})$ and
$(\rho_{C},\rho_{D})$ on either side of each are approximately the
values that would arise at coexistence of the two given phases, in the absence
of the third. These pairs of values of $\rho$ may be found from the bulk free
energy density by the usual double-tangent construction, as shown in figure
\ref{binodals} (which also shows the construction for the globally stable
binodal values $(\rho_{\alpha},\rho_{\beta})$ for \mbox{1--3} phase
coexistence). In principle the double tangent construction is subject to small
corrections due to interface motion; these are
calculated, for completeness, in Appendix \ref{correction} but we neglect them
here.

\subsection{Growth or Collapse?}
The time evolution of such an interface is not strictly a question of
steady-state (or even quasi-steady-state) dynamics. Accordingly we give only a
brief discussion, and leave a fuller exploration of this interesting problem to
the companion paper \cite{Evans97b} (see also \cite{Evanslet}). The basic issue
is whether the slab of metastable phase grows or shrinks.

\begin{figure}
  \epsfxsize=11cm
  \begin{center}
  \leavevmode\epsffile{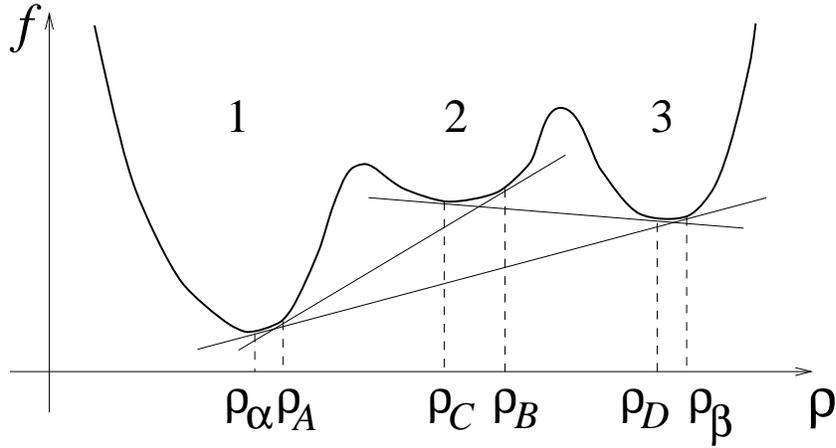}
  \caption{Figure demonstrating the double-tangent construction to find
   the stable binodal densities $(\rho_{\alpha},\rho_{\beta})$ and the
   metastable binodal densities $(\rho_{A},\rho_{B})$ and
   $(\rho_{C},\rho_{D})$ in a three-well potential. This potential can be
   converted to the form of fig.\ref{3wells} by adding or subtracting a linear
   term ($f \to f + a\rho + b$), and then both shifting the origin of $\rho$,
   and rescaling it.}
  \label{binodals}
  \end{center}
\end{figure}

When curvature is
small, as is the case {\em between} the interfaces, the Cahn-Hilliard equation
is well approximated by the diffusion equation, $\dot{\rho}=D\nabla^2\rho$, with
the diffusivity given by $D=\Gamma f_{2}(\rho)$ which is approximately
constant given that
$\rho$ does not vary much. ($D$ is exactly constant in a quadratic
potential well of fixed $\Gamma$.) Notice in figure \ref{binodals} that the
inequality
$\rho_{B}>\rho_{C}$ is a {\it necessary result} of the metastability of well
2. Hence, since the diffusion equation governs the inter-wall region,
$j_{2}$ is positive. Thus flux flows onto the \mbox{2--3} interface,
contributing positively to $v_{2}$ and negatively to $v_{1}$.
So, the effect of
$j_{2}$ is to {\em reduce} $\Delta x$, as would be expected for the
dynamics of a
metastable phase. If phase 2 is to grow, the constant flux $j_{1}$ into the
system must be sufficiently large to make $|v_{1}|>|v_{2}|$. Invoking the
diffusion approximation, and the linearity of the function $\rho(x)$ in
phase 2, the condition for growth of the metastable phase (rather than
recombination of the \mbox{1--3} interface) becomes
\begin{equation}
\label{criterion}
  j_{1} > D \frac{(\rho_{B}-\rho_{C})}{\Delta x} \left(
             1+\frac{\rho_{B}-\rho_{A}}{\rho_{D}-\rho_{C}}\right).
\end{equation}

When this condition is satisfied, a `split' mode of interfacial propagation can
arise, which is fundamentally different from the propagation of an unsplit
\mbox{1--3} interface in a number of respects. The most important distinction is
that now, if $j_1$ is held constant, the width of the metastable region,
$\Delta x$ grows without limit. In contrast, in a stable \mbox{1--3} interface,
$\Delta x$ remains bounded. Indeed, in the equilibrium ($v=0$) interface,
$\Delta x$ is of the order of the characteristic interfacial width
$\sim\sqrt{K/f_{2}}$, and at higher speeds, $\Delta x$ becomes smaller.
Accordingly there is an upper bound on $\Delta x$ close to the equilibrium
value (although not equal to it, since unsplit propagation is resumed after a
small, positive perturbation in $\Delta x$).

Another qualitative
difference between the split and unsplit modes of propagation is their
response to a perturbation. It was demonstrated in Section
\ref{stability} that increasing the width $\Delta x$ of the metastable
region led to an increase in the flux though it, resulting in a
{\em negative} feedback mechanism. On the other hand, if the interface is
split, and therefore non-monotonic and containing a well-developed region in
which diffusive motion is dominant over curvature-induced motion, increasing
$\Delta x$ reduces the gradient in region 2. This reduces $j_{2}$, and
causes the \mbox{2--3} interface to lag still further behind the
\mbox{1--2} wall. So the corresponding feedback in split interface motion is
{\em positive}. It follows that, at constant $j_1$, there is a barrier (in
configuration space) to the formation of an unsplit interface, but once this
barrier is crossed, such an interface will remain split indefinitely.

\subsection{Selection of Split or Unsplit Mode}
\label{selection}
It has been shown that (at least for piecewise-quadratic
potentials) the propagation of an unsplit interface is locally stable,
and that, given sufficient input flux, the split interface
mode is also ``stable" (in the sense of remaining split indefinitely). The
question arises of which mode of evolution will be selected in a given system,
and how it might be possible to change from one to the other. Clearly in a real
system, the flux input to an interface is not constant. Normally, in late-stage
evolution, it is a diminishing function of time. One might conclude from this
that the criterion for growth of the split mode (equation \ref{criterion})
must at some point be violated. However, this criterion becomes easier
to satisfy as $\Delta x$ increases.
The ultimate fate of a split interface in fact depends on supersaturation: this
is described in the companion paper
\cite{Evans97b}.

The problem of how to `unbind' a \mbox{1--3}
interface is somewhat clearer: a large transient increase in input flux
is required, to overcome the  negative feedback mechanism described in Section
\ref{stability}. There is presumably some critical value of $\Delta x$ at which
the feedback switches from negative to positive and the interface
splits. The transient increase in flux must be sufficient to separate
the leading (\mbox{1--2}) part of the interface from the trailing part
by this critical amount, {\em before} the interface can deliver a
restorative increase in flux to region 3 by curvature-induced motion.
If the transient increase of input flux is insufficient, and the
\mbox{1--3} interface adjusts to the new higher speed, the criterion
for unbinding it becomes more stringent (since higher-speed interfaces
are narrower, and hence both further from the critical value of $\Delta
x$, and `stiffer' in terms of the negative feedback mechanism). Since
fluxes tend to decrease with time during late-stage phase-ordering, a
transient increase in flux, sufficient to unbind a \mbox{1--3}
interface, is most likely to occur during the early
stage dynamics (nucleation or spinodal decomposition). These dynamics are
not quasi-steady-state, and we do not discuss them further here. But it
is interesting that, whenever metastable phases are possible, the {\em details}
of these early stage dynamics can determine the gross features (split vs.
unsplit
mode) of phase separation at much later times.

Finally, in the context of mode selection, a useful distinction can be drawn
between two types of interfacial binding/unbinding. The type described above can
be called ``curvature unbinding" -- the process whereby $\Delta x$ becomes large
compared to $\sqrt{K/f_2}$. In the companion paper
\cite{Evans97b} (see also \cite{Evanslet}) we study ``diffusive unbinding" which
is linked to the evolution of $j_1$ (treated above as an externally imposed
parameter). In Ref.\ \cite{Evanslet}, we also report numerical results which
show that curvature unbinding certainly does occur, within the
Cahn-Hilliard equation, at least for some parameter values and some
initial conditions. These include cases (such as an initial step function
wall between phases 1 and 3) which, though strongly perturbed from the
equilibrium profile, are definitely not unbound to begin with.

\section{Conclusion}
\label{conclusion}

It is often assumed that steady-state
solutions of the Cahn-Hilliard equation (Model B), for the phase
ordering dynamics of a conserved order parameter, are unphysical and
therefore uninteresting. In this article we have shown that
such solutions shed light on  the local dynamics of interfaces, so long as care
is taken to interpret the boundary conditions correctly. Exact solutions were
found for the case of any piecewise-quadratic order parameter potential and
piecewise-constant mobility; these are locally stable. For the most
general case of arbitrary bulk free energy density and mobility as
functions of mass density $\rho$, a systematic expansion scheme was derived
(Appendix \ref{generic}) to solve Model B for a moving interface.
Using this scheme, it can be shown (to at least first order in $v$)
that an interface contracts when moving under condensation,
and expands during evaporation. If a metastable phase
exists, whose density is intermediate to the two stable phases, a
moving interface between the stable phases is still locally stable with
respect to small perturbations at all speeds, but may be `split' by a
large, transient increase in the flux of condensing material, and thereafter
exhibit a qualitatively different mode of propagation. Such a transient
disturbance is most likely to arise in the early stages of nucleation
and growth. The split mode of interface motion results in the formation
of a macroscopically large amount of the metastable phase and relies on a
sufficiently large flux of condensing material being maintained.
A companion paper \cite{Evans97b} (see also
\cite{Evanslet}), discusses further the implications of these findings
for the growth rates of competing stable and metastable domains during the
phase-ordering process.

\section{Acknowledgements}

This work was supported by EPSRC Grant No.\ GR/K56025. We thank Wilson Poon
for a series of illuminating discussions.

\appendix

\section{Exact Solution of the Steady-State Equation for
Piecewise-Quadratic Potential}
\label{solution}

  Equation \ref{steady} is now solved to find $\rho(s)$ for any
piecewise-quadratic potential $f(\rho)$ and piecewise-constant mobility
$\Gamma(\rho)$. Between
discontinuities in $f_{1}$, $f_{2}$ or $\Gamma$, $\rho(s)$ solves the equation
\begin{displaymath}
  \rho''''-a\rho''-b\rho'=0
\end{displaymath}
where $a=f_{2}/K$ and $b=v/(K\Gamma)$. The solution is
\begin{displaymath}
  \rho'(s) = \sum_{j=1}^3 A_{j} e^{\omega_{j}s}
\end{displaymath}
for some constants $A_{j}$, fixed by the boundary conditions and matching
conditions. The constants $\omega_{j}$ are the three roots of the cubic
equation
\begin{displaymath}
  \omega_{j}^3-a\omega_{j}-b=0
\end{displaymath}
which is solved, for $27b^2<4a^3$, by
\begin{eqnarray*}
  \omega_{1} &=& 2 \sqrt{\frac{a}{3}} \sin\left(\frac{\sigma-2\pi}{3}\right) \\
  \omega_{2} &=& 2 \sqrt{\frac{a}{3}} \sin\left(\frac{\sigma}{3}\right)  \\
  \omega_{3} &=& 2 \sqrt{\frac{a}{3}} \sin\left(\frac{\sigma+2\pi}{3}\right)
\end{eqnarray*}
where
\begin{displaymath}
  \sigma \equiv -\arcsin\sqrt{\frac{27b^2}{4a^3}}.
\end{displaymath}
If $27b^2>4a^3$, the solutions may be written in terms of the real quantities
\linebreak
\mbox{$
\alpha\equiv(\frac{1}{2}b+\frac{1}{2}\sqrt{b^2-\frac{4}{27}a^3})^{1/3}
$} and \mbox{$
\beta\equiv(\frac{1}{2}b-\frac{1}{2}\sqrt{b^2-\frac{4}{27}a^3})^{1/3}
$} as
\begin{eqnarray*}
  \omega_{1} &=& \alpha+\beta  \\
  \omega_{2} &=& -\mbox{$\frac{1}{2}$}(\alpha+\beta)
    + i \mbox{$\frac{\sqrt{3}}{2}$}(\alpha-\beta) \\
  \omega_{3} &=& \omega_{2}^*
\end{eqnarray*}
All that remains to be found is the vector of coefficients
$
  \mbox{\boldmath$A$}\equiv\left(       \begin{array}{c}
                                          A_{1} \\ A_{2} \\ A_{3}
                                        \end{array}                     \right)
$
for each piece. This is fixed by assigning values to $\rho'$,
$\rho''$ and $\rho'''$ at the right-hand boundary of each section, which may
be re-defined as the origin of $s$ by multiplying {\boldmath$A$} by an
appropriate exponential factor. From the equation for $\rho'(s)$, it follows
that
\begin{displaymath}
  \mbox{\boldmath$A$} = \left(
  \begin{array}{ccc}
        1               &       1               &       1               \\
        \omega_{1}      &       \omega_{2}      &       \omega_{3}      \\
        \omega_{1}^2    &       \omega_{2}^2    &       \omega_{3}^2
  \end{array}
  \right)^{-1} \left(
  \begin{array}{c}
        \rho'(0)        \\
        \rho''(0)       \\
        \rho'''(0)
  \end{array}
  \right).
\end{displaymath}
The vector of derivatives of $\rho$ is determined, from the solution in
the neighbouring piece, by the matching conditions given
in Section \ref{exact}, where it was stated that the gradient, chemical
potential and flux must all be continuous. That is, the quantities
$\rho'$, $(K\rho''-f_{1})$ and $(K\rho'''-f_{2}\rho')\Gamma$ are
continuous. These quantities can be calculated from the neighbouring
solution, given the value of $s$ at which it meets the discontinuity
(say $s=s_{1}$). If the discontinuity (in the mobility or potential) is
at a value $\rho=\rho_{1}$, then $s_{1}$ is given by inverting the
equation $\rho(s_{1})=\rho_{1}$ which, unfortunately cannot in general
be done analytically. (However, the inversion is always possible for the
reasons explained in Section \ref{nosplit}.) Hence one numerical step is
required in the solution. (Of course the special case of a double-well
potential, with just one cusp discontinuity, is completely soluble analytically,
since the arbitrary origin of
$s$ can be put at the cusp of
$f(\rho)$.)

  Finally, the boundary condition that $\rho\rightarrow\rho_{\infty}$ as
$s\rightarrow+\infty$ results in the solution for the highest density section:
$
  \mbox{\boldmath$A$} = \left(  \begin{array}{c}
                                -(\rho_{\infty}-\rho_{1})\:\omega_{3} \\
                                                            0   \\
                                                            0
                                        \end{array}   \right)
$
and hence \newline
\mbox{$\rho'(s)=-(\rho_{\infty}-\rho_{1})\:\omega_{3}\:e^{\omega_{3}s}$}
for $s>0$, and $\rho(0)=\rho_{1}$.

\section{Integral Solution of Cahn-Hilliard}
\label{generic}

In Section \ref{properties}, the assertion was made that interfaces
generically contract when moving under condensation. This result and
its converse (that interfaces expand during evaporation) will be
derived now, using a systematic expansion of steady-state
solutions of the Cahn-Hilliard equation in powers of the interface
velocity $v$. As previously, we consider the one-dimensional
 Cahn-Hilliard equation in a frame moving at velocity $v$, in which
distance is measured by the coordinate $s$; the order
parameter $\rho$ asymptotes to a finite constant $\rho_{\infty}$ as
$s\rightarrow+\infty$. However, in contrast to the analysis in Section
\ref{exact} and Appendix \ref{solution}, no particular form will be
assumed for the order parameter potential
$f(\rho)$, and likewise the mobility $\Gamma(\rho)$ will be an
arbitrary function.

\subsection{Inversion of Variables}

 It is convenient to invert the equation so that
$\rho$ becomes the independent variable, and the equation is solved for
$s$. This means that the unspecified potential $f(\rho)$ and mobility
$\Gamma(\rho)$ are now functions of the {\em independent} variable.
Henceforth let the curvature constant $K$ be set to unity without loss
of generality. (This is equivalent to measuring time in units of $K$
and length in units of $\surd K$.) After inversion, the full,
one-dimensional Cahn-Hilliard equation in a moving frame may be written
\begin{displaymath}
  \frac{\partial s}{\partial t} \,\, = \,\, \frac{\partial}{\partial\rho} \left[
    \left(\frac{\partial s}{\partial\rho}\right)^{\!-1}\!\! \Gamma \:
    \frac{\partial^2}{\partial\rho^2} \left\{ \frac{1}{2}\left(
    \frac{\partial s}{\partial\rho}\right)^{\!-2}\!\!-f \right\} \right] -v.
\end{displaymath}
Clearly (as is well-known) adding a linear term to
the potential $f(\rho)$ has no effect.

\subsection{Integration of the Steady-State Equation}

  Let us now set the time derivative to zero in this moving frame, and
integrate once with respect to $\rho$, to obtain the steady-state
equation
\begin{displaymath}
  \Gamma(\rho) \left( \frac{ds}{d\rho} \right)^{\!-1} \frac{d^2}{d\rho^2}
    \left\{ \frac{1}{2}\left(\frac{ds}{d\rho}\right)^{\!-2}\!\!
    -f(\rho)+\lambda\rho+c \right\}\,\, = \,\, (\rho-\rho_{\infty}) v
\end{displaymath}
where the arbitrary linear term $-(\lambda\rho+c)$ has been explicitly
added to $f(\rho)$. Note that the constant $\lambda$ is equal to the
chemical potential in the asymptotically flat region: $\lambda =\mu(\infty)$
(remember $\mu$ is non-uniform in a moving interface). It
can be confirmed that $-v\rho_{\infty}$ is the correct constant of
integration, given that the left-hand side of the above equation is
simply the flux $j$.

  Let us define $h(\rho)\equiv ds/d\rho$, so that
$1/h=\nabla\rho$. Then integrating with respect to $\rho$, we find
\begin{displaymath}
  \frac{d}{d\rho}(\mbox{$\frac{1}{2}$} h^{-2}-f+\lambda\rho+c)
   = v \int \frac{(\rho-\rho_{\infty})h}{\Gamma}\,d\rho+\mbox{const.}
\end{displaymath}
It is easy to confirm that the left-hand side of this equation is the
chemical potential, measured with respect to the value at
$\rho_{\infty}$. Now consider the factors in the integrand. As
$\rho\rightarrow\rho_{\infty}$, $(\rho-\rho_{\infty})$ tends to zero
linearly, while it is expected (and confirmed in the special case of
appendix \ref{solution}) that $h(\rho)\rightarrow\infty$
logarithmically. Hence the integrand vanishes as
$\rho\rightarrow\rho_{\infty}$. So the constant in the above equation
is zero, and we have
\begin{displaymath}
  \frac{d}{d\rho}(\mbox{$\frac{1}{2}$} h^{-2}-f+\lambda\rho+c)
    = -v \int_{\rho}^{\rho_{\infty}}
    \frac{(\rho'-\rho_{\infty})}{\Gamma(\rho')}h(\rho')\,d\rho'.
\end{displaymath}
Let us henceforth absorb the terms $\lambda\rho+c$ into the definition
of $f(\rho)$. (That is, the arbitrary linear part of $f(\rho)$ is
defined so that the potential has zero value and gradient at
$\rho_{\infty}$.) Integrating by parts once gives the final result
\begin{equation}
\label{integral}
  h(\rho) = \frac{1}{\sqrt{2f(\rho)}} \left\{ 1
     - \frac{v}{f(\rho)}\int_{\rho}^{\rho_{\infty}}
    \frac{(\rho_{\infty}-\rho')(\rho'-\rho)}{\Gamma(\rho')}
    h(\rho')\,d\rho' \right\}^{-\frac{1}{2}}
\end{equation}
Notice that the Cahn-Hilliard equation, in its differential form, is
{\em fourth} order, but that this integral representation of the steady-state
equation contains only {\em one}
integration.

  If $h(\rho)$ is written as a power-series in $v$, then the binomial
in equation \ref{integral} may be expanded, and the formula iterated to
produce a systematic series approximation for $h$, to arbitrarily high
powers of the velocity. The power-series expansion of the steady-state
solution $s(\rho)$ is then obtained by integration as:
\begin{displaymath}
  s(\rho) = \int h(\rho)\,d\rho.
\end{displaymath}
Clearly the origin of $s$ (and hence the constant of integration) is
arbitrary.

\subsection{First-Order Correction to Interfacial Width}

  This method is now applied to find the first-order correction to the
width $\Delta s$ of a moving interface, defined as the distance between
two points on the interface at which the densities have certain fixed
values $\rho_{1}$ and $\rho_{2}$ (which could be the densities of the
two maxima in a three-well potential, for example). It transpires that
\begin{eqnarray*}
  \Delta s &=& \Delta s_{0} + c v + {\cal O}(v^2)  \\
  \mbox{where} \,\, \Delta s_{0} &=& \int_{\rho_{1}}^{\rho_{2}}
    \frac{d\rho}{\sqrt{2f(\rho)}}  \\
  \mbox{and} \,\,\,\, c &=& \frac{1}{4}\int_{\rho_{1}}^{\rho_{2}} d\rho\,
    f(\rho)^{-\frac{3}{2}}
    \int_{\rho}^{\rho_{\infty}} d\rho'\,(\rho_{\infty}-\rho')(\rho'-\rho)
    \Gamma(\rho')^{-1} f(\rho')^{-\frac{1}{2}}.
\end{eqnarray*}
This last equation expresses the rate of change of width of the
interface with velocity, $c$ in terms only of the two functions which
characterize the physics of the system; the bulk free energy density
$f(\rho)$ and the mobility $\Gamma(\rho)$. All factors in the
integrands of this expression are positive over the ranges of
integration, so $c$ is positive. Negative values of $v$ correspond to
growth of the asymptotically flat, dense region by condensation, and
positive values correspond to evaporation. So, without assigning any
special properties to the functions $f$ and $\Gamma$, it has been shown
that the interface contracts ($\Delta s<\Delta s_{0}$) during
condensation and expands during evaporation.

  As mentioned at the end of Section \ref{properties}, $\rho_{\infty}$
may vary from its equilibrium value ({\em i.e.\ }the density at which
phase 3 coexists in equilibrium with phase 1 for a three-well
potential). In the present section, a formalism has been developed by
which the solution for a moving interface is expanded about the $v=0$
solution. Does this allow for variation in $\rho_{\infty}$? The answer
is `yes' because there is in fact a whole family of $v=0$ solutions
with different values of $\rho_{\infty}$, of which the equilibrium
solution is just one member. The equilibrium solution is the special
member of this family for which $\rho$ asymptotes to a finite constant
as $s\rightarrow-\infty$, rather than growing exponentially (either
positive or negative) and thus remaining curved and having unphysical
boundary conditions at $s\rightarrow-\infty$. But, for our purposes,
the whole family may be used since, as discussed in Section
\ref{exact}, the left-hand boundary is not put at negative infinity.

\section{Correction to the Double-Tangent Construction for a Moving Interface}
\label{correction}

  The double-tangent construction illustrated in figure \ref{binodals}
gives the {\em equilibrium} densities of two coexisting phases. Recall
its elementary derivation as follows \cite{DeHoff93}. In a {\em uniform}
part of the system of volume $V$, containing $N$ particles, the local
density is $\rho=N/V$, and the free energy is $F=fV$. From these two
simple relations, and the definitions of chemical potential $\mu=(\partial
F/\partial N)_{V}$ and pressure $P=-(\partial F/\partial V)_{N}$, it follows
that $f=\mu\rho-P$, which is the equation of a straight line on a plot of
$f$ {\em versus} $\rho$, with gradient $\mu=f'(\rho)$ and intercept
$-P$. Given that two coexisting phases have equal chemical potentials
and pressures, it follows that the straight lines tangent to $f(\rho)$
at the respective coexistent densities have equal gradients and
intercepts. Hence they are the same line.

  Consider now the non-equilibrium case of an interface in uniform
motion at velocity $v$. By continuity, the flux at any point on the
interface is $j=-(\rho_{\infty}-\rho)v$. In model B,
$\nabla\mu=-j/\Gamma$. Hence, integrating across the interface, the
difference in chemical potential between the dense, asymptotically
uniform phase at positive infinity, and any given point $A$ is
\begin{displaymath}
  \Delta\mu \equiv \mu(\infty)-\mu(A)
   = v\int_{A}^{\infty}\frac{(\rho_{\infty}-\rho)}{\Gamma}dx
\end{displaymath}
from which the equilibrium result follows for $v=0$.

  Now let us consider the local quantity $P\equiv\mu\rho-f$ (which reduces
to the usual definition of pressure in a homogeneous system). With this
definition,
\begin{displaymath}
  \nabla P = \rho\nabla\mu - K\nabla^2\rho\nabla\rho.
\end{displaymath}
Using again the above expression for $\nabla\mu$, and integrating gives
\begin{displaymath}
  \Delta P \equiv P(\infty)-P(A)
    = \left[-\frac{1}{2}K\left(
    \frac{\partial\rho}{\partial x}\right)^2\right]_{A}^{\infty}
    -\int_{A}^{\infty}\frac{\rho j}{\Gamma}dx.
\end{displaymath}
If the curvature part of the chemical potential is small at point $A$,
this expression for the pressure difference becomes
\begin{displaymath}
  \Delta P =-\mbox{$\frac{1}{2}$}
   K\left[\left(\frac{j}{\Gamma f''}\right)^2\right]_{A}^{\infty}
    -\int_{A}^{\infty}\frac{\rho j}{\Gamma}dx
    = -\int_{A}^{\infty}\frac{\rho j}{\Gamma}dx + {\cal O}(v^2).
\end{displaymath}
So the equality of chemical potentials and pressures across a static
interface, which gives rise to the double-tangent construction, has the
following corrections, to first order in $v$, for a moving interface:
\begin{eqnarray*}
  \Delta\mu &=& v\int_{A}^{\infty}\frac{(\rho_{\infty}-\rho)}{\Gamma}\:dx  \\
  \Delta P &\approx&
     v\int_{A}^{\infty}\frac{(\rho_{\infty}-\rho)\:\rho}{\Gamma}\:dx.
\end{eqnarray*}
Notice that both corrections depend on the range of integration, {\em
i.e.\ }on the position of the point $A$. This is no surprise, since
there must be gradients in the pressure and chemical potential outside
the interface in order induce motion. Notice also that both
coefficients of $v$ are positive, so $\Delta\mu$ and $\Delta P$ are of
the same sign as $v$. This is negative for condensation and positive
for evaporation. So, as expected, during condensation the pressure
difference and chemical potential difference across an interface are
smaller than at equilibrium.

\end{document}